\documentclass[prb,12pt]{revtex4}
\begin{document}

\title{On the origin of the C induced $p4g$ reconstruction of Ni(001)}
\author{Sergey Stolbov, Sampyo Hong, Abdelkader Kara, and Talat S. Rahman}%
\affiliation{Department of Physics, Kansas State University, 116 Cardwell Hall, Manhattan, Kansas 66506, USA}%
\begin{abstract}
First principles calculations of the geometric and electronic
structures have been performed for two coverages (0.25 ML and 0.5
ML) of C on Ni(001) to understand the mechanism of the Ni(001)
reconstruction induced by carbon adsorption. The calculated
structural behavior of the system is in a good agreement with
experimental observations. The calculated path and energetics of
the $c(2\times 2)$ -- $p4g$ reconstruction in C$_{0.5}$/Ni(001) is
provided. A dramatic reduction of the local electronic charge on
adsorbed carbon is found to occur upon the reconstruction that
decreases the electron-electron repulsion on C site. This effect
together with the formation of covalent bonds between C and the
second layer Ni atoms, leads to reconstruction of Ni(001).
\end{abstract}
\maketitle
\section{Introduction}
Nickel is used as a catalyst for many reactions. The presence of
carbon in the reaction environment may lead to atomic adsorption
on the metal surface changing the electronic structure and
consequently its reactivity \cite{sto04}. It is thus important to
understand the mechanism by which C adsorbs on Ni surfaces. On the
other hand, this system is an excellent prototype for addressing
fundamental issues such as the effects of the adsorption of light
elements on the electronic and geometric structures of metal
substrate and the character of the chemical bonding between the
adsorbate and the surface. It is thus not surprising that the
first observation of C induced reconstruction of Ni(001)
\cite{onu79} has attracted much attention of experimentalists and
theorists who aim to provide increasingly accurate description and
explanation of the nature of this phenomenon
\cite{kli93,onu79,ter00,gau91,rah85,rah87,sun99,sun00,mul86,san92,kir03,alf99,hon04}.

It is well known that when carbon adsorbs on Ni(001) with coverage
less than one third of a monolayer (ML) it occupies hollow sites
and does not change the symmetry of the outmost metal layers
\cite{kli93}. As the coverage exceeds 0.33ML, the surface
reconstructs to what has become known as "the clock
reconstruction" \cite{kli93,onu79,ter00}. Fig. 1  illustrates
schematically these two geometries, in both of which C atoms
occupy the fourfold hollow sites. In the reconstructed structure,
the topmost Ni atoms are displaced by alternate clock-wise and
counter-clock-wise rotation around C, forming a geometry of the
$(2\times 2)p4g$ symmetry. A large volume of additional
information has also been obtained from more recent experiments.
Scanning tunneling microscopy (STM) \cite{kli93} and photoelectron
diffraction (PhD) \cite{ter00} measurements have been performed
for non-reconstructed Ni(001) with the C coverage less than 0.3ML
(C$_{\theta}$/Ni(001), $\theta < 0.3$ML). Although the
quantitative results in the two papers differ, both report a small
outward radial displacement of the topmost Ni atoms (Ni1) around
the C adsorbate. On the other hand, for $\theta > 0.33$ML, results
of low energy electron diffraction (LEED) \cite{gau91} and PhD
\cite{ter00} experiments reveal a large (0.4 -- 0.55\AA) lateral
displacement of the Ni1 atoms and penetration of C atoms into the
Ni1 layer that indicate the surface reconstruction.

The nature of this reconstruction has been a subject of much
debate in the past two decades. Using a simple force constant
model in lattice dynamical calculations. Rahman et al.
\cite{rah85,rah87} have suggested that the reconstruction is
driven by phonon softening. Several scenarios for the reduction of
specific force constants were presented, including those involving
coupling of C to the Ni atoms in the second layer (Ni2)
\cite{kli93,onu79, rah87,mul86} which seem plausible since C
penetrates the surface during the reconstruction. On the other
hand, Terborg et al. \cite{ter00} argued that the reconstruction
is caused  by C-induced Ni1-Ni1 repulsion rather than by C-Ni2
bonding, a conclusion that has been upheld in the recent band
structure calculations \cite{kir03}. Charge transfer in the system
has also been suggested to be a factor controlling the
reconstruction \cite{sun99,sun00}. However, to our knowledge, the
distribution of local charges in the C/Ni(001) system have not yet
been calculated from first principles. It has also been proposed
that surface stress \cite{mul86,san92} might be responsible for
the reconstruction. {\it Ab initio} calculations by Alfe et al.
\cite{alf99}, however emphasize chemical bonding effects rather
than elastic effects in causing the reconstruction. Direct first
principles calculations of C induced surface stress on Ni(001)
\cite{hon04} also do not reveal any specific relationship between
surface stress and surface reconstruction. Both Alfe et al.
\cite{alf99} and Hong et al. \cite{hon04} have addressed the issue
of geometric and electronic structural changes brought about by
the adsorption of 0.5 ML of C on Ni(001). Both used the plane wave
pseudopotential (PWPP)  methods with the local density
approximation for the exchange-correlation potential. The authors
of Ref. \onlinecite{alf99} compare the local densities of the
electronic states (LDOS) for the non-reconstructed and
reconstructed structures. However, no distinct tractable
modification of LDOS is found upon the reconstruction. In Ref.
\onlinecite{hon04} the valence charge density for the
non-reconstructed and reconstructed structures is provided, but
authors suggest that a more detailed study of the charge density
is required.

The nature of the reconstruction of Ni(001) thus remains to be
unclear, and its understanding requires detailed description of
the electronic structure, character of chemical bonding and
energetics of the system. Since the reconstruction occurs at a
certain C coverage $\theta \approx 0.33$ ML, it is also important
to compare and contrast the electronic structures of
C$_{\theta}$/Ni(001) with $\theta < 0.33$ ML and $\theta > 0.33$
ML.

In this paper, we present the results of our calculations of the
geometric and electronic structure of C/Ni(001) performed for the
0.25 ML and 0.5 ML C coverages within the density functional
theory with the generalized gradient approximation for the
exchange-correlation potential. Applying a structural optimization
procedure within the PWPP method, we reproduce the reconstruction
for C$_{0.5}$/Ni(001) and provide the energetics and path of
transition from $c(2\times 2)$ to $p4g$ geometry. The valence
charge densities and local valence charges are calculated for the
systems under consideration using more accurate full-potential
linearized augmented plain wave method. Analyzing modification of
these characteristics upon change in the C coverage and the
reconstruction we provide a consistent description of the
mechanism of the reconstruction.

\section{Computational Details}

All calculations presented in this paper have been performed
within the density functional theory with the generalized gradient
approximation (GGA) for the exchange-correlation functional
\cite{per92}. The Ni(001) surface was modelled by a tetragonal
supercell consisting of a 7 layer Ni slab and 11 \AA \/ of vacuum.
For both 0.25 ML and 0.5 ML carbon coverages we have used
(2$\times$2) two dimensional unit cell which included four Ni
atoms per layer. The supercell has also contained two C atoms (one
atom at each side of the Ni slab) for the 0.25 ML coverage and
four C atoms (two atoms at each side of the slab) for the 0.5 ML
coverage. Thus, the supercells have consisted of 30 and 32 atoms
for 0.25 ML and 0.5 ML coverages, respectively.

Surface geometry and energetics have initially been calculated
using the PWPP method \cite{pay92} with ultrasoft pseudopotentials
\cite{van90} used for C and Ni. We set cutoff energies for the
plane-wave expansion of 300 eV for both clean and C-covered
surfaces and used a 5x5x1 Monkhorst-Pack $k$-point mesh in the
Brillouin zone sampling \cite{mon76}. The structures were
optimized until the forces acting on each atom converged better
than 0.01 eV/\AA.

These optimized structures were then used as input for a more
detailed analysis of the electronic structure, which includes
self-consistent calculations of the valence charge density and
local charges of the systems, using the full-potential linearized
augmented plane wave (FLAPW) method \cite{sin94} as embodied in
the computational code Wien2k \cite{bla01}. Optimized structures
obtained from PWPP calculations were used as input data for Wien2k
which further refined the geometries after a few ionic iterations.
be consistent, we have used the same supercells and k-point mesh
in the Brillouin zone for the FLAPW calculations as in the PWPP
ones. In the FLAPW method, the local charges are calculated
through integration over muffin-tin (MT) spheres of radius
$R_{MT}$ centered at each atom. To analyze the effect of the
adsorbate on these specific quantities the set of $R_{MT}$ should
be chosen to be the same for all structures under consideration.
Ideally, $R_{MT}$ should be as large as possible without causing
the MT spheres to overlap. For inner atoms of the Ni slab,
including the Ni2 atoms, a choice of $R_{MT}=1.19$\AA\/ is
optimal. However, for Ni1, $R_{MT}=1.005$\AA\/ provided more
compatibility which the shorter C -- Ni bond lengths. For C atoms
$R_{MT}=0.773$\AA\/ was needed. To keep plane wave cutoff high
enough ($RK_{max}$=7) with the reduced $R_{MT}$ 's for the surface
atoms, we used basis sets of about 5500 LAPW 's including 256
local orbitals for the surface with 0.25 ML coverage and for the
reconstructed surface with 0.5 ML coverage and 2800 LAPW's
including 128 local orbitals for the non-reconstructed surface
with 0.5 ML coverage.

\section{Results and Discussion}
\subsection{Carbon induced structural modification of Ni(001)}

\subsubsection{Effects of 0.25 ML C on Ni(001) geometric structure}

To our knowledge, the case of 0.25 ML C on Ni(001) has not been
studied so far using first principles calculations based on
density functional theory. W3e have carried out such calculations
of the geometric structure of the system using a structural
optimization procedure that minimizes the forces applied to the
atoms. We find our results to be in a good agreement with the
experimental observations. Namely, the structural optimization
does not lead to the p4g reconstruction in this system and results
in a stable c(2$\times$2) geometry. Even if we initially shift the
Ni1 atoms along the p4g reconstruction path, they revert to the
c(2$\times$2) geometry during the optimization. We find that the
presence of C causes radial outward displacement $\Delta r =
0.07$\AA\/ of the Ni1 atoms from C. This number is in the range of
the experimental results, which spread from $\Delta r = 0.02 \pm
0.03$\AA\/ (for 0.15 ML coverage) \cite{ter00} to $\Delta r = 0.15
\pm 0.15 $\AA (for about 0.2 ML coverage)\cite{kli93}. The
calculated equilibrium C-Ni1 separation $d_{01}=0.24$\AA\/ is
pretty close to $d_{01}=0.21 \pm 0.07$\AA\/ obtained for
C$_{0.15}$/Ni(001) from PhD measurements \cite{ter00}.

\subsubsection{Effects of 0.5 ML C on Ni(001) geometric structure}

We starts our calculation of C$_{0.5}$/Ni(001) with the perfect
c(2$\times$2) structure, and calculate the total energy
($E_{tot}$) of the system as a function of $d_{01}$. At each step
we fix the C position and let the rest of atoms relax (move any
direction). The result of these calculations is shown on the right
side of Fig. 2. No reconstruction occurs during structural
optimization and a $E_{tot}$ minimum is found at $d_{01}=0.53$\AA.
Next, we repeat the procedure, but starting with the surface Ni
atoms (Ni1) initially in the p4g reconstructed geometry
\cite{ter00}. Since the nickel atoms are free to relax as the C
atoms are lowered, the system finds the stable position of Ni1. We
find that although Ni1 positions change during optimization
depending on $d_{01}$, the system is kept in the p4g structure.

Since Ni is a ferromagnetic metal, we perform these calculations
for both non-magnetic and spin-polarized ferromagnetic cases. The
values of $d_{01}$ and the lateral Ni1 displacement $\Delta xy$
obtained for the equilibrium (minimum $E_{tot}$) structures from
these two calculations are listed in Tab. \ref{tab:dxy-p4g}
together with those measured in experiments. As seen from the
table, the spin polarization effect on the geometric structure is
negligible and both calculated structures are in very good
agreement with the experimental observations. To understand why
the magnetic effect on the geometric structure is so small, we
calculate local Ni spins in the system by integrating spin density
over the spheres of 1.2\AA\/ radius centered at different Ni
atoms. We find that while the spin on the Ni atoms located at the
center of the slab equals $0.69 \mu_B$, it is reduced to $0.39
\mu_B$ on Ni2 neighboring carbon and diminished to $0.02 \mu_B$ on
Ni1. This result indicates that strong covalent C-Ni bonding
suppresses spin polarization on the Ni atoms surrounding adsorbed
C and makes it possible to neglect the magnetic effects on the
C/Ni(001) surface geometric structure. We thus perform all
following calculations for non-magnetic systems.

The $E_{tot}(d_{01})$ dependence calculated for non-magnetic p4g
structure is plotted on the left side of Fig. 2. One can see that
the $E_{tot}$ minimum for the p4g structure is about 0.8 eV lower
than that for the c(2$\times$2) one. At first sight, both
dependences shown in Fig. 2 suggest that the system has two energy
minima: the local one for the c(2$\times$2) structure and the
global one for the reconstructed p4g structure and there is an
energy barrier between them. However, our further consideration
shows that this is a wrong conclusion. C adsorbed in the ideal
c(2$\times$2) structure cannot induce forces acting on Ni atoms
along any direction, which breaks the c(2$\times$2) symmetry.
Regular optimization procedures applying the quasi-Newton or
conjugate gradient methods are thus not able to lead such a system
to reconstruction. In reality, even if the system appears to be in
the c(2$\times$2) structure, atomic vibrations naturally break the
symmetry that can cause the forces leading to the reconstruction.

To place these speculations on firmer ground, we perform another
calculation starting with a system slightly deviated from the
(2$\times$2) symmetry. Namely we start with the quasi-equilibrium
c(2$\times$2) structure obtained above, but with Ni1 atoms
slightly (by 0.01\AA) shifted along the reconstruction path and
let all atoms relax. We find that the structural optimization
leads C$_{0.5}$/Ni(001) to the equilibrium p4g structure with
parameters ($d_{01}$ and $\Delta xy$) practically the same as
obtained in the simulation described above. The system thus
undergoes a c(2$\times$2) -- p4g transformation without any energy
barrier that is in agreement with the conclusion made in Ref.
\cite{alf99}. The path of the Ni1 and C displacements and the
changes in $E_{tot}$ obtained during the optimization are shown in
Fig. 3. One can see that the Ni1 lateral displacement and C normal
shift happen simultaneously, although first Ni1 move faster than C
, then, after $\Delta xy$ reaches the value of about 0.4\AA, it
changes slow, while $d_{01}$ sharply decreases by 0.3\AA. At the
first stage of the reconstruction the total energy of the system
substantially decreases, while at the second stage
$E_{tot}(d_{01})$ is flat.

\subsection{Electronic Structure of C/Ni(001)}

Now, when we find that the structural behavior of C/Ni(001)
obtained from our calculations completely reproduces the
experimental observations, we are going to analyze the
relationship between the geometric and electronic structures of
the system to reveal factors controlling the clock reconstruction.
To this end, using FLAPW method, we calculate the electronic
structure for two observed systems: $c(2\times
2)$C$_{0.25}$/Ni(001) and $p4g$C$_{0.5}$/Ni(001), as well as, for
two hypothetical $c(2\times 2)$C$_{0.5}$/Ni(001) structures: one
with quasi-stable geometry ($d_{01}=0.53$\AA\/) and another with a
reduced C-Ni1 separation ($d_{01}=0.38$\AA). The valence charge
densities calculated for these structures are plotted along a
plane perpendicular to the surface in Figs. 4 -- 7. We find strong
covalent C-Ni1 bonds to be formed in all four considered
structures. These bonds are seen in the figures as "bridges" of
electronic density between the C and Ni1 atoms. The $c(2\times
2)$C$_{0.25}$/Ni(001) and $p4g$C$_{0.5}$/Ni(001) structures also
have distinct "bridges" between the C and Ni2 atoms, whereas both
$c(2\times 2)$C$_{0.5}$/Ni(001) structures do not. We find thus
that the stable structures observed in experiments differ from
unstable ones by the presence of extra C-Ni2 covalent bonds. The
extra covalent bond naturally reduces the total energy of the
system and increases its stability.

Now the question emerges why the C-Ni2 bonds are not formed in
$c(2\times 2)$C$_{0.5}$/Ni(001). This bond can occur if the C-Ni2
bond length ($l_{C-Ni2}$) is short enough. In Tab. \ref{tab:bl} we
provide values of $l_{C-Ni2}$ as well as, the C-Ni1 bond lengths
($l_{C-Ni1}$) calculated for the four considered structures. As
seen from the table, in the stable $p4g$C$_{0.5}$/Ni(001) and
$c(2\times 2)$C$_{0.25}$/Ni(001), in which distinct C-Ni2 covalent
bonds are formed, $l_{C-Ni2}$ are much shorter than in the
unstable structures. However, as we reduce $l_{C-Ni2}$ in
$c(2\times 2)$C$_{0.5}$/Ni(001) from 2.33\AA\/ to 2.18\AA, the
total energy of the system increases (see Fig. 2). To understand
the cause of such a behavior, we focus on local charges on carbon.
The local charges $Q_{MT}(C)$, calculated by integrating the
valence charge density over the C muffin-tin sphere with
$R_{MT}=0.773$\AA, are listed in Tab. \ref{tab:q-phi}. We find
that if $l_{C-Ni2}$ in $c(2\times 2)$C$_{0.5}$/Ni(001) is reduced
by 0.25\AA, $Q_{MT}(C)$ substantially increases. This is caused by
reduction of $l_{C-Ni1}$ that takes place when $l_{C-Ni2}$
decreases: as $l_{C-Ni1}$ becomes shorter, neighboring Ni1 atoms
compress the electronic charge density on carbon. This should
enhance the electron-electron repulsion at the C atom that may
cause the increase in the total energy mentioned above. The
reduction of $l_{C-Ni1}$ happens, because in $c(2\times
2)$C$_{0.5}$/Ni(001) there is no room for Ni1 outward
displacement. As the clock reconstruction happens, $l_{C-Ni1}$
increases and $Q_{MT}(C)$ dramatically decreases. It should be
noted that $Q_{MT}(C)$ in stable $c(2\times 2)$C$_{0.25}$/Ni(001)
is also significantly smaller than in the unstable $c(2\times 2)$
structures with the 0.5 ML coverage. However, the C-Ni1 bond
length is not the only factor controlling $Q_{MT}(C)$: if the
C-Ni2 covalent bond is formed some portion of the valence electron
density is transferred to the C-Ni2 covalent "bridge". This is
seen from comparison of $Q_{MT}(C)$ in $c(2\times
2)$C$_{0.25}$/Ni(001), $p4g$C$_{0.5}$/Ni(001) and $c(2\times
2)$C$_{0.5}$/Ni(001) with $d_{01}=0.53$\AA. All three structures
have the same $l_{C-Ni1}=1.83$\AA\/ (see Tab. \ref{tab:bl}), but
different $Q_{MT}(C)$ (see Tab. \ref{tab:q-phi}). The C-Ni2
covalent bond is not formed in $c(2\times 2)$C$_{0.5}$/Ni(001),
and $Q_{MT}(C)$ in this system is the largest. On the other hand,
$p4g$C$_{0.5}$/Ni(001) has the strongest C-Ni2 covalent bond and
the lowest value of $Q_{MT}(C)$. We find thus two factors that
make the $c(2\times 2)$ -- $p4g$ reconstruction favorable in
C$_{0.5}$/Ni(001): formation of C-Ni2 covalent bond that reduces
the energy of the system and reduction of $Q_{MT}(C)$ that
prevents strong electron-electron repulsion on carbon.

Since interaction of dipoles formed in the surface upon adsorption
could also affect surface geometry \cite{sun00}, we calculate the
work functions $\Phi$ of the systems which reflect change in the
dipole moments. We find, however, that $\Phi$ correlates rather
with the C-Ni1 separation than with in-plane displacements of the
Ni1 atoms.

In summary, we have performed first principles calculations of the
geometric and electronic structures of C/Ni(001) with the C
coverages of 0.25 and 0.5 ML. The structural behavior of the
system obtained from the calculations is in a good agreement with
experimental observations. No energy barrier is found on the
reconstruction path. We find that formation of the C-Ni2 covalent
bonds is crucial for stability of C/Ni(001). These bonds are
present in the stable $c(2\times 2)$C$_{0.25}$/Ni(001) and
$p4g$C$_{0.5}$/Ni(001) structures and absent in unstable
$c(2\times 2)$C$_{0.5}$/Ni(001). Apart from formation C-Ni2
covalent bond, the $c(2\times 2)$ -- $p4g$ reconstruction in
C$_{0.5}$/Ni(001) is accompanied by a dramatic reduction of the
local electronic charge on adsorbed carbon that is supposed to
decrease the electron-electron repulsion on C site.

\begin{acknowledgements}
Grant from NCSA, Urbana, IL was beneficial in providing
computational resources. We acknowledge financial support from
NSF, USA, under grant No. CHE-0205064.
\end{acknowledgements}

\begin{table}
\caption{\label{tab:dxy-p4g} Comparison of structural parameters
of C$_{0.5}$-p4g/Ni(001) calculated in this work with those
obtained from experiments \cite{gau91,ter00}. NM and FM denote
non-magnetic and ferromagnetic calculations, respectively}
\begin{ruledtabular}
\begin{tabular}{ccc}
 & $d_{01}$(\AA) & $\Delta xy$(\AA)  \\
\hline
 NM  & 0.13 & 0.51  \\
 FM  & 0.14 & 0.51 \\
 Exper. & 0.1--0.12 & 0.41 -- 0.55 \\
\end{tabular}
\end{ruledtabular}
\end{table}
\begin{table}
\caption{\label{tab:bl} C-Ni bond lengths calculated for the
considered C/Ni(001) structures.}
\begin{ruledtabular}
\begin{tabular}{ccc}
 & $l_{C-Ni1}$ (\AA) &  $l_{C-Ni2}$ (\AA) \\
\hline
$p4g$-C$_{0.5}$    & 1.83 & 1.99 \\
$c(2\times 2)$-C$_{0.25}$ & 1.83 & 2.04 \\
$c(2\times 2)$-C$_{0.5}, d_{01}=0.38$ & 1.79 & 2.18 \\
$c(2\times 2)$-C$_{0.5}, d_{01}=0.53$ & 1.83 & 2.33 \\
\end{tabular}
\end{ruledtabular}
\end{table}
\begin{table}
\caption{\label{tab:q-phi} C local charges and work functions
calculated for the considered C/Ni(001) structures.}
\begin{ruledtabular}
\begin{tabular}{ccc}
 & $Q_{MT}$(C) & $\Phi$(eV)  \\
\hline
$c(2\times 2$-C$_{0.5}, d_{01}=0.53$\AA & 4.58 & 5.67  \\
$c(2\times 2)$-C$_{0.5}, d_{01}=0.38$\AA & 4.67 & 5.55 \\
$p4g$-C$_{0.5}, d_{01}=0.13$\AA  & 4.30 & 5.39 \\
$c(2\times 2)$-C$_{0.25}$ & 4.41 & 5.35 \\
\end{tabular}
\end{ruledtabular}
\end{table}

\section{Figure captions}
Fig. 1 (Color online). Illustration of the $c(2\times
2)$C$_{0.25}$/Ni(001) (upper panel) and $p4g$C$_{0.5}$/Ni(001)
(lower panel) geometric
structures. \\

Fig. 2. Dependence of the total energy versus the C-Ni1 separation
calculated for C$_{0.5}$/Ni(001) in $c(2\times 2)$ and $p4g$
geometries. Zero energy corresponds to C desorption.\\

Fig. 3. Reconstruction path and energetics calculated for
C$_{0.5}$/Ni(001).  \\

Fig. 4 (Color online). Valence charge density calculated for
$c(2\times 2)$C$_{0.25}$/Ni(001) and plotted along the plane
perpendicular to
the surface. \\

Fig. 5 (Color online). The same as in Fig. 4, but for $p4g$C$_{0.5}$/Ni(001). \\

Fig. 6 (Color online). The same as in Fig. 4, but for $c(2\times
2)$C$_{0.5}$/Ni(001) with
$d_{01}=0.53$\AA. \\

Fig. 7 (Color online). The same as in Fig. 4, but for $c(2\times
2)$C$_{0.5}$/Ni(001) with $d_{01}=0.38$\AA.

\end{document}